\documentclass[twocolumn,amsmath,amssymb,floatfix,prl,footinbib,superscriptaddress]{revtex4-1}

\usepackage{amsmath,amssymb,natbib,bm,graphicx,url,epsf,color}
\usepackage[english]{babel}
\usepackage{wasysym}
\usepackage{MnSymbol}
\usepackage{amsmath}
\usepackage{graphicx}
\usepackage{pdfpages}
\usepackage[colorinlistoftodos]{todonotes}
\usepackage[colorlinks=true, allcolors=blue]{hyperref}

\newcommand{\VDS}{V_{\mathrm{DS}}}
\newcommand{\Vds}{V_{\mathrm{DS}}}
\newcommand{\Vg}{V_{\mathrm{g}}}

\makeatletter
\AtBeginDocument{\let\LS@rot\@undefined}
\makeatother

\begin{document}
\title{Strongly correlated charge transport in silicon MOSFET quantum dots}

\author{M. Seo}
\affiliation{SPEC, CEA, CNRS, Universit\'e Paris-Saclay, CEA Saclay, 91191 Gif-sur-Yvette cedex, France
}
\author{P. Roulleau}
\affiliation{SPEC, CEA, CNRS, Universit\'e Paris-Saclay, CEA Saclay, 91191 Gif-sur-Yvette cedex, France
}
\author{P. Roche}
\affiliation{SPEC, CEA, CNRS, Universit\'e Paris-Saclay, CEA Saclay, 91191 Gif-sur-Yvette cedex, France
}
\author{D.C. Glattli}
\affiliation{SPEC, CEA, CNRS, Universit\'e Paris-Saclay, CEA Saclay, 91191 Gif-sur-Yvette cedex, France
}
\author{M. Sanquer}
\affiliation{Univ. Grenoble Alpes, CEA, INAC-PHELIQS, 38000 Grenoble, France
}
\author{X. Jehl}
\affiliation{Univ. Grenoble Alpes, CEA, INAC-PHELIQS, 38000 Grenoble, France
}
\author{L. Hutin}
\affiliation{CEA, LETI, Minatec Campus, 38000 Grenoble, France
}
\author{S. Barraud}
\affiliation{CEA, LETI, Minatec Campus, 38000 Grenoble, France
}
\author{F.D. Parmentier}
\affiliation{SPEC, CEA, CNRS, Universit\'e Paris-Saclay, CEA Saclay, 91191 Gif-sur-Yvette cedex, France
}

\date{\today}

\begin{abstract}

Quantum shot noise probes the dynamics of charge transfers through a quantum conductor, reflecting whether quasiparticles flow across the conductor in a steady stream, or in syncopated bursts. We have performed high-sensitivity shot noise measurements in a quantum dot obtained in a silicon metal-oxide-semiconductor field-effect transistor. The quality of our device allows us to precisely associate the different transport regimes and their statistics with the internal state of the quantum dot. In particular, we report on large current fluctuations in the inelastic cotunneling regime, corresponding to different highly-correlated, non-Markovian charge transfer processes. We have also observed unusually large current fluctuations at low energy in the elastic cotunneling regime, the origin of which remains to be fully investigated.
\end{abstract}

\maketitle


Current fluctuations in a mesocopic conductor, or quantum shot noise \cite{Blanter2000}, reflect the granularity of charge transfers across the conductor. By measuring low frequency current fluctuations, one can probe the correlations between subsequent charge transfers and quantify how random these transfers are. Those correlations are underpinned by the interplay between the quantum statistics of the particles flowing across a given conductor, electronic interactions, and the physical mechanisms giving rise to transport in the conductor. Rare, uncorrelated charge transfers lead to Poissonian shot noise $S_{II}=2e \langle I \rangle$, with $e$ the charge of the quasiparticles and $\langle I \rangle$ the average value of the dc current flowing across the conductor. Correlations between subsequent transfers are encoded in the \textit{Fano factor} $F$, defined as the ratio between the shot noise and its Poissonian value. Fermionic statistics tend to impose some order on charge transport \cite{Reznikov1995, Kumar1996} , characterized by reduced fluctuations ($F<1$), or, in the case of perfectly ballistic conductors, fully noiseless transport ($F=0$). While Coulomb interactions tend to do the same \cite{Birk1995}, they can, in some remarkable cases, give rise to positively correlated transport processes with super-Poissonian ($F>1$) fluctuations, where charges flow in bursts through the conductor. 

Coulomb-blockaded quantum dots are very rich systems, as they can not only present Poissonian and sub-Poissonian transport regimes \cite{Birk1995, Blanter2000}, but also, depending on their internal structure, strongly correlated transport with super-Poissonian current fluctuations \cite{Sukhorukov2001, Belzig2005, Thielmann2005a, Onac2006, Weymann2007, Aghassi2008, Carmi2012, Kaasbjerg2015, Zhang2007, Zarchin2007, Fricke2007, Okazaki2013, Harabula2018}. The latter regime corresponds to non-Markovian transport processes where the transfer of a charge across the dot changes its state, thereby influencing the next transfer event \cite{Sukhorukov2001}. As a result, the quantum dot randomly switches between highly and poorly conducting channels while other parameters (\textit{e.g.} temperature, bias voltage) remain fixed. This can happen in the sequential tunneling regime if several levels of a dot, with markedly different couplings to the leads, participate to transport. In that case the current across the quantum dot shows random telegraph signal features \cite{Fricke2007}, yielding strongly enhanced fluctuations \cite{Belzig2005, Zhang2007, Zarchin2007, Fricke2007, Harabula2018}.

\begin{figure}[ht]
\centering
\includegraphics[width=0.45\textwidth]{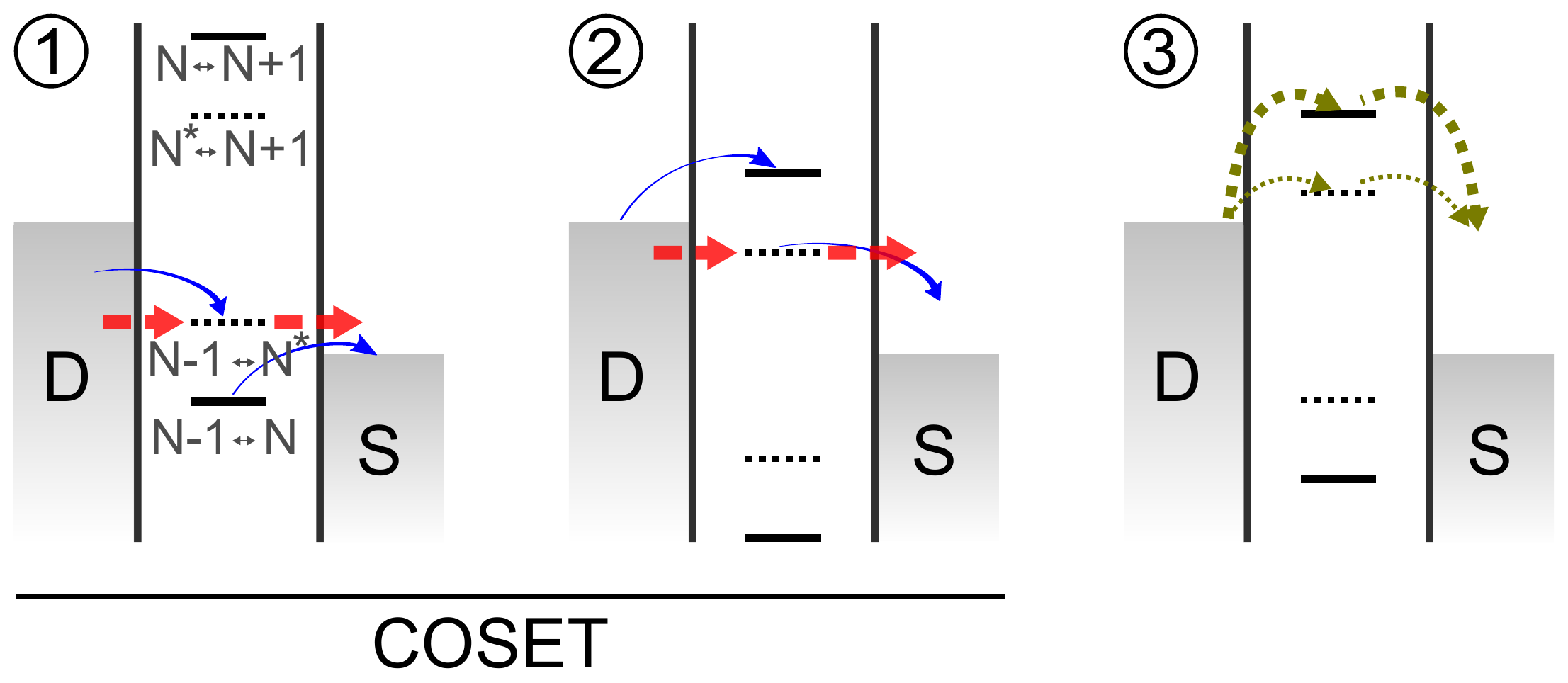}
\caption{\label{fig1} Sketches of cotunneling processes giving rise to super-Poissonian noise, described in the text. The thin blue arrows correspond to usual inelastic cotunneling; the thick, dashed, red arrows to COSET. In \raisebox{.5pt}{\textcircled{\raisebox{-.9pt} {3}}}, the dashed green arrows correspond to elastic cotunneling with different rates depending on the state of the dot.}
\end{figure}

Inelastic cotunneling \cite{DeFranceschi2001} is also expected to enhance the current fluctuations, as it is, by definition, a charge transfer process that changes the state of the quantum dot \cite{Sukhorukov2001}. Several mechanisms leading to super-Poissonian shot noise in the inelastic cotunneling regime have been proposed, depending on the internal structure of the quantum dot and the respective chemical potentials of the dot and the leads. Some of these mechanisms are depicted in Fig.~\ref{fig1}, for a $N$-charge quantum dot with a single excited state labeled $N^\star$. Inelastic cotunneling (blue arrows) events leave the dot in the excited state $N^\star$. In \raisebox{.5pt}{\textcircled{\raisebox{-.9pt} {1}}} (resp. \raisebox{.5pt}{\textcircled{\raisebox{-.9pt} {2}}}), the $N-1\leftrightarrow N^\star$ (resp. $N^\star\leftrightarrow N+1$) transition involving $N^\star$ sits in the bias window, allowing direct transport through the dot after the cotunneling event \cite{Wegewijs2001, Schleser2005}. This process, usually referred to as cotunneling-assisted tunneling (CAST \cite{Golovach2004, Aghassi2008}, or COSET \cite{Andergassen2010, Kaasbjerg2015, Gaudenzi2017, Harabula2018}), is depicted as dashed red arrows in Fig.~\ref{fig1}. It leads to enhanced fluctuations, as the system randomly switches between the blocked state $N$ and the conducting (excited) state $N^\star$ \cite{Thielmann2005a, Weymann2007, Aghassi2008, Kaasbjerg2015}. In \raisebox{.5pt}{\textcircled{\raisebox{-.9pt} {3}}}, all transitions are outside the bias window, and the dot is always in a blocked state. This can nonetheless lead to enhanced fluctuations: indeed, whether the quantum dot is blocked in the ground or excited state, \textit{elastic} cotunneling (green dashed arrows in Fig.~\ref{fig1}) with \textit{a priori} different rates can occur. Inelastic cotunneling events will then randomly switch the quantum dot between states with different (albeit small) conductances, yielding super-Poissonian fluctuations \cite{Aghassi2008, Kaasbjerg2015}. This latter regime occurs if $\Delta^* < E_c/3$, where $\Delta^*$ is the excited state energy and $E_c$ the charging energy \cite{Wegewijs2001, Golovach2004, Aghassi2008, Kaasbjerg2015, Gaudenzi2017, Harabula2018}. Previous experimental works reported the observation of super-Poissonian noise in carbon nanotubes \cite{Onac2006, Harabula2018} and in GaAs/AlGaAs \cite{Okazaki2013} quantum dots in the inelastic cotunneling regime. In refs. \cite{Okazaki2013, Harabula2018}, the structure of the excited states suggest that the enhanced shot noise is due to COSET processes; however, no clear modulations in the Fano factor, such as the ones predicted in refs. \cite{Aghassi2008, Kaasbjerg2015} that discriminate between the mechanisms described above, have been observed so far. In this letter, we present the first clear observation of such modulations.

\begin{figure}[ht]
\centering
\includegraphics[width=0.35\textwidth]{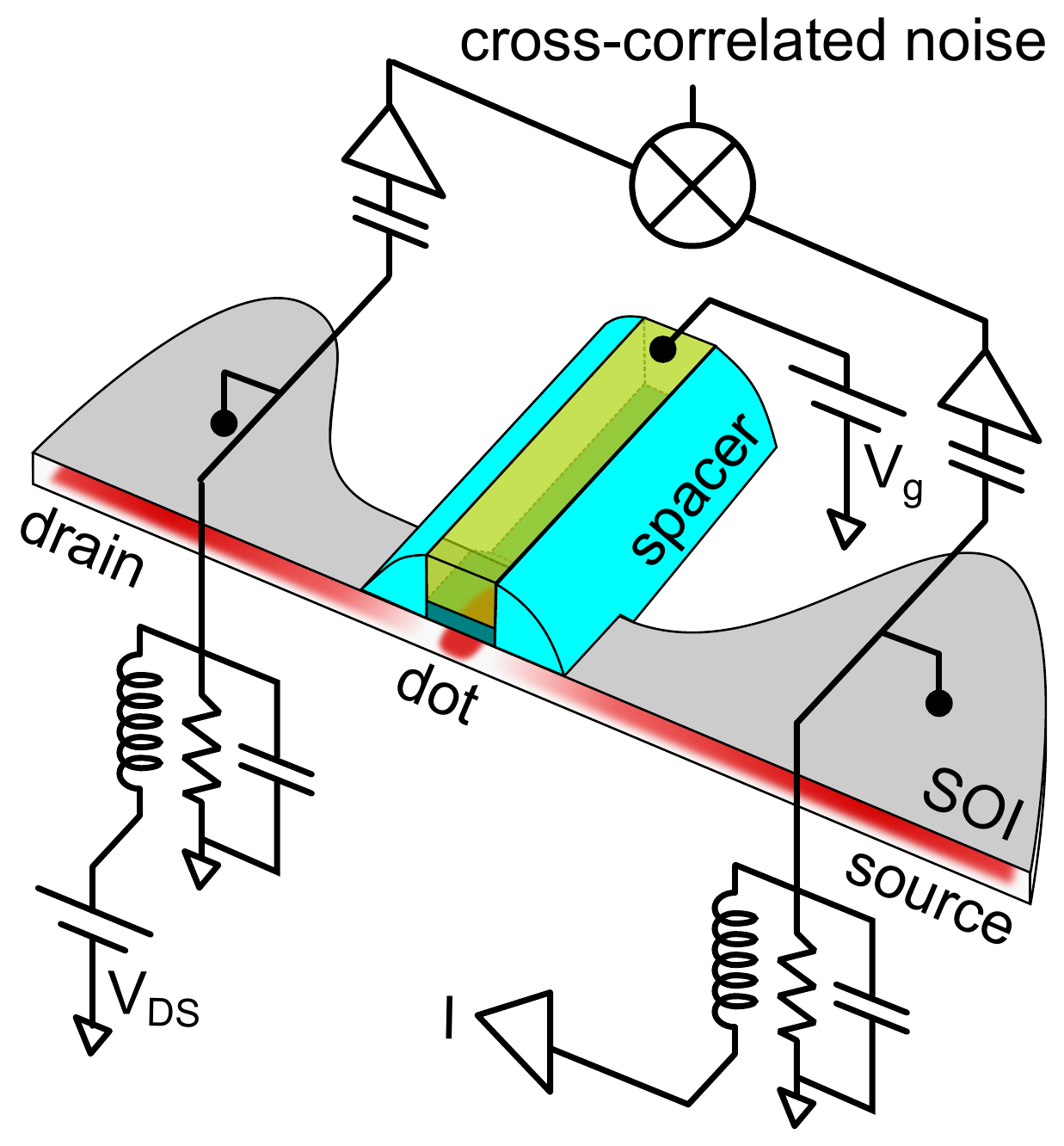}
\caption{\label{fig2-sample}Description of the experiment. The charge carriers in the silicon on insulator (SOI) channel are symbolized by the red clouds, showing the drain and source reservoirs on either side of the quantum dot. The spacers, realizing the tunnel barriers coupling the dot to the leads, are shown in light blue. The gate is shown in yellow and the gate insulator in dark grey. The electrical schematic is a simplified description of the measurement setup, including the pair of $RLC$ tanks used in the shot noise measurements.}
\end{figure}

We have investigated quantum dots formed in silicon nanowire metal-oxide-semiconductor field-effect transistors (MOSFETs), fabricated using a  microelectronics technology based on 300~ mm silicon-on-insulator (SOI) wafers. The low-temperature electronic conductance properties of such devices have been extensively studied in previous works \cite{Sanquer2000, Boehm2005, Hofheinz2006, Hofheinz2006a, Pierre2010, Prati2012, Roche2012, Lavieville2015, Voisin2016}, showing very robust Coulomb blockade characteristics. We performed conductance, current and  shot noise measurements in several small-size ($\approx 20\times30\times 10$~nm) p-type devices using the setup described in Fig.~\ref{fig2-sample}, at a temperature of $0.3$~K in a cryogen-free He$-3$ refrigerator. We measure the excess shot noise $S_{II}$ in the quantum dot with a cross-correlation technique \cite{diCarlo2006, Parmentier2016,SM}, where the current fluctuations on either side of the device are filtered at low temperature by $RLC$ tanks with a resonance frequency of approx.~$3.5$~MHz, and amplified using home-made ultra-low noise preamplifiers. The cross-correlation of the outputs of both preamplifiers is then computed using a high speed digitizer. The dc and low frequency signals through the quantum dot (drain-source voltage $\VDS$, dc current $I$, and differential conductance $dI/d\VDS$) are applied and measured through the inductor of each $RLC$ tanks. This allows us to simultaneously measure $S_{II}$ and $dI/d\VDS$ as a function of $\VDS$ and gate voltage $\Vg$. In the most stable devices, we can compare the simultaneously acquired differential conductance and shot noise with the current, obtained in separate acquisitions, with very small drifts in gate voltage over extended periods of time (typically $2.5$~mV in a week). From the independents measurements of $S_{II}$ and $I$, we compute an effective Fano factor $F=S_{II}/2eI$.

\begin{figure}[ht]
\centering
\includegraphics[width=0.49\textwidth]{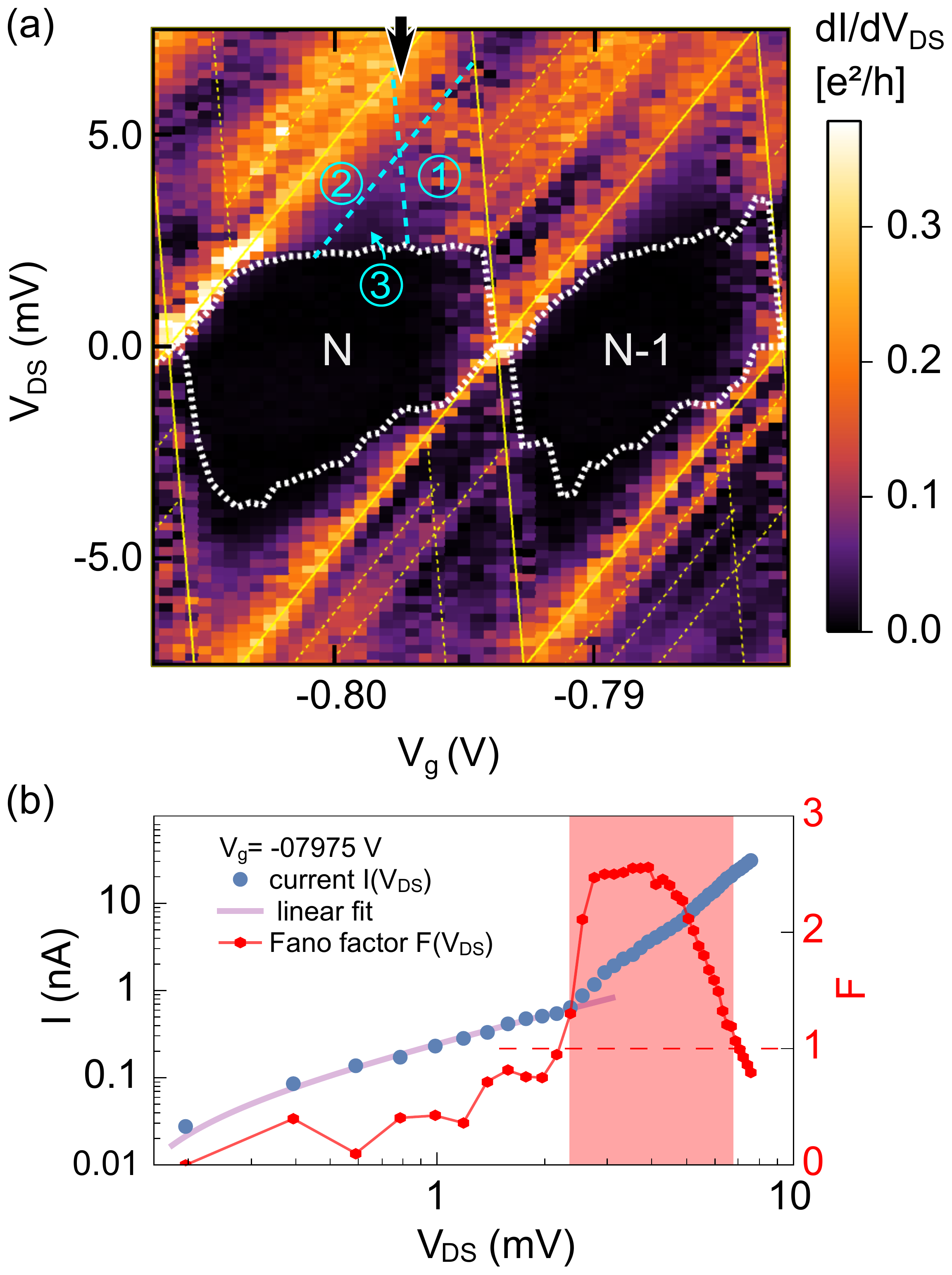}
\caption{\label{fig3-GandI}\textbf{(a)} Differential conductance $dI/d\VDS$ measured as a function of gate voltage $\Vg$ and drain-source voltage $\VDS$. The continuous yellow lines locate the edges of the Coulomb diamonds, and the dashed yellow lines the excited states. The thick dotted white lines show the onset of inelastic cotunneling, extracted from current measurements (see below). The regions labeled \raisebox{.5pt}{\textcircled{\raisebox{-.9pt} {1}}}, \raisebox{.5pt}{\textcircled{\raisebox{-.9pt} {2}}} and \raisebox{.5pt}{\textcircled{\raisebox{-.9pt} {3}}}, separated by the light blue dashed lines (see text for their construction), correspond to the processes depicted in Fig.~\ref{fig1}. \textbf{(b)} Current $I$ (blue symbols, left axis) and Fano factor $F$ (red symbols, right axis) as a function of $\VDS$ in (resp.) log-log and semi-log scale, for $\Vg=-0.7975~$V (black vertical arrow in \textbf{(a)}). Continuous violet line: linear fit of $I(\VDS)$ at low $\VDS$ \cite{SM}. The horizontal red dashed line indicates the Poisson value $F=1$. The red area shows the inelastic cotunneling range, delimited by the value of $\Vds$ at which $I(\VDS)$ deviates from a linear behavior, and the by edge of the Coulomb diamond.
}
\end{figure}


Figure~\ref{fig3-GandI}a shows measurements of $dI/d\VDS$ in our best device, displaying two Coulomb diamonds (identified by the number of holes $N-1$ and $N$), analyzed in this letter (measurements over the full range of $\Vg$, as well as in other devices, are shown in the supplementary materials \cite{SM}). The edges of the diamonds indicate the transitions involving ground states of the quantum dot, highlighted by the continuous yellow lines in Fig.~\ref{fig3-GandI}a. Similarly, the transitions involving excited states of the quantum dot, associated to resonances in $dI/d\VDS$ outside of the diamonds, are highlighted by the dashed yellow lines. The negative slope resonance lines, corresponding to quantum dot transitions aligned with the electrochemical potential of the source, are constructed by comparing $dI/d\Vds$ and $I(\VDS)$ \cite{SM}. We extract the lever arms \cite{SM}, as well as the charging and typical excited states energies $E_c\approx8.7$~meV, $\Delta^* \approx2.2$~meV. Note that these parameters are not constant over the full range of $\Vg$, as the shape of the quantum dot is modified for large excursions in gate voltage \cite{SM}. $dI/d\VDS$ is non-zero inside the diamonds, indicating the presence of cotunelling processes. The aforementioned condition $\Delta^* < E_c/3$ to observe the various cotunneling regimes \raisebox{.5pt}{\textcircled{\raisebox{-.9pt} {1}}}, \raisebox{.5pt}{\textcircled{\raisebox{-.9pt} {2}}} and \raisebox{.5pt}{\textcircled{\raisebox{-.9pt} {3}}} is fulfilled in our device. This is shown by extending the (yellow dashed) excited transitions lines into the Coulomb diamonds, yielding the light blue dashed lines in Fig.~\ref{fig3-GandI}a. As demonstrated below, these regimes indeed give rise to different shot noise contributions. 

We quantitatively discriminate elastic and inelastic cotunneling regimes by exploiting our measurements of the current $I$ flowing across the dot. We check that the measured current \cite{SM} shows $\VDS$ and $\Vg$ dependences accurately matched by the resonance lines shown in Fig.~\ref{fig3-GandI}a. We then extract the onset of inelastic cotunneling, plotted as thick, dotted white lines in Fig.~\ref{fig3-GandI}a, by tracking, for a given value of $\Vg$, at which $\VDS$ does $I$ change from a linear $\VDS$ dependence (characteristic of \textit{elastic} cotunneling) to a power-law dependence (characterizing \textit{inelastic} cotunneling) \cite{Averin1990, Glattli1991}. The onset closely follows the edges of the diamonds in vicinity of the charge degeneracy points, then clearly bifurcate inside the diamonds upon reaching the first excited state resonance. Furthermore, $dI/d\VDS$ is clearly non-zero beyond the onset lines. Note that the onset lines do not correspond to constant $\VDS$, which is reminiscent of energy renormalization due to inelastic cotunneling \cite{Holm2008, Splettstoesser2012}. The extraction of the onset is illustrated in Fig.~\ref{fig3-GandI}b for $\Vg=-0.7975$~V, plotting $I(\VDS)$ in log-log scale (left Y-axis, blue dots). Below $\VDS\approx2.3$~mV, defining the onset, $I(\VDS)$ is linear (the violet line is a linear fit of the data in that range), then takes a power law $\left( \VDS \right)^\alpha$, with $\alpha\approx~3-7$ (red area) \cite{SM}. Plotting the Fano factor $F$ on the same graph (right Y-axis, red symbols) shows the noise contribution of each cotunneling regime: while $F$ stays below the Poissonian value $F=1$ in the elastic cotunneling regime, it rapidly shoots up to significantly large values $F\approx~2.5$ in the inelastic cotunneling regime, then decreases back to sub-Poissonian values outside of the Coulomb diamond.


\begin{figure}[ht]
\centering
\includegraphics[width=0.48\textwidth]{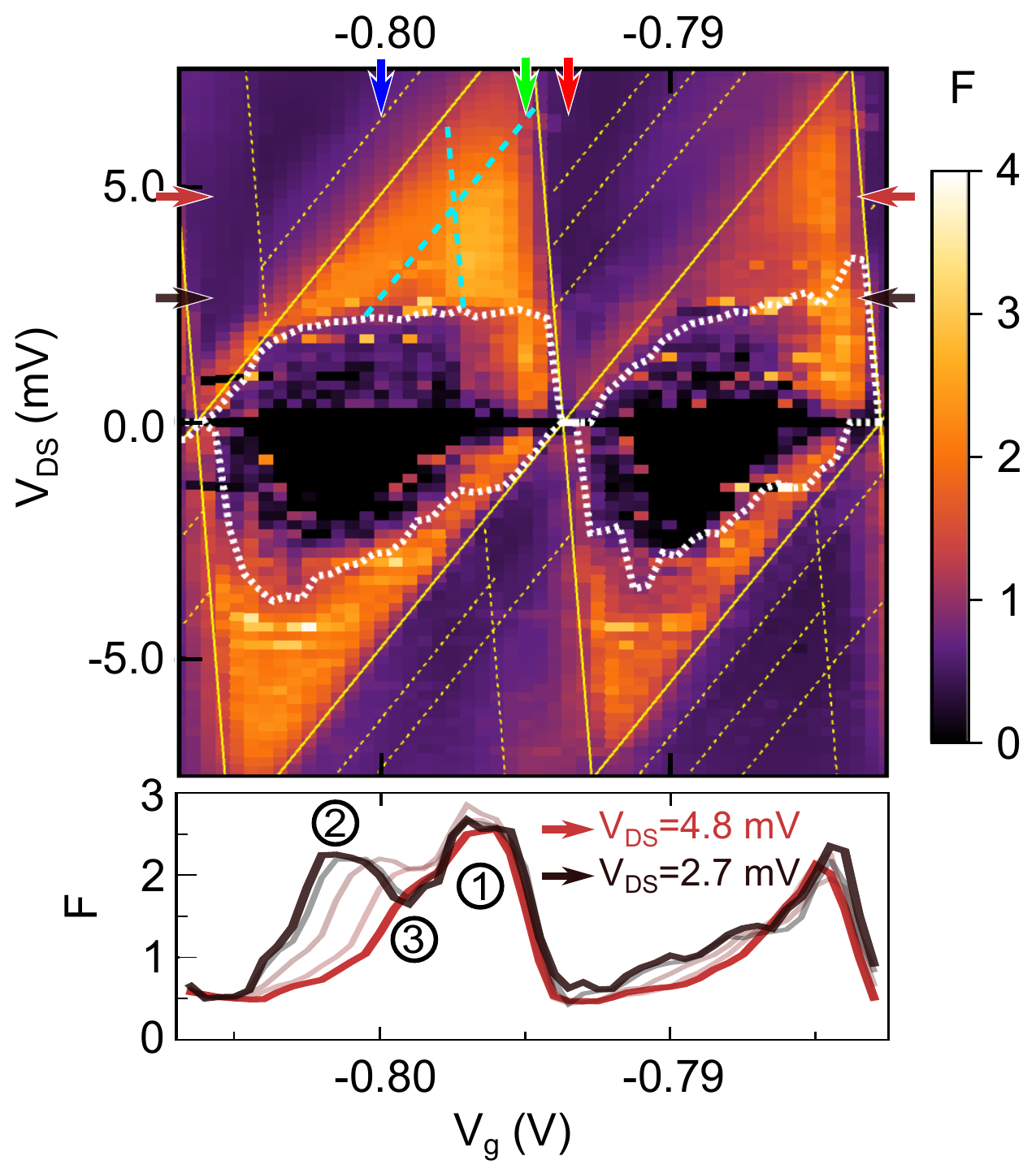}
\caption{\label{fig4}Top:~Fano factor $F$ measured as a function of $\Vg$ and $\VDS$. The (full and dashed) yellow lines, the dashed blue lines and the white dashed lines are the same as in Fig.~\ref{fig3-GandI}. The blue, green and red vertical arrows indicate the line cuts at constant $\Vg$ shown in Fig.~\ref{fig5}. The brown and dark brown horizontal arrows indicate the line cuts at constant $\VDS$ shown in the bottom panel.  Bottom:~Line cuts of $F$ as a function of $\Vg$, for $\VDS$ between $2.7$~(dark brown) and $4.8$~mV (brown); the translucent lines correspond to intermediary values of $\VDS$.}
\end{figure}

Fig.~\ref{fig4} shows a map of $F$ over the same range of $\VDS$ and $\Vg$ as in Fig.~\ref{fig3-GandI}a, also including the resonance lines and the inelastic cotunneling onset presented in the latter. With a few notable exceptions, which we discuss below, $F$ is, very clearly, only above the Poisson value in the regime of inelastic cotunneling. In the elastic cotunneling regime, $F$ generally takes values smaller than $1$. Note that given the small magnitude of both $I$ and $S_{II}$ in this regime, $F$ presents large relative fluctuations. For clarity, we have set $F=0$ whenever either $I<50$~pA or $S_{II}<2\times10^{-29}$~A$^2/$Hz. Outside of the diamonds, $F$ is close to $0.5$, the value for sequential tunneling across a dot with symmetric barriers \cite{Blanter2000}.

We now discuss the variations of $F$ in regions \raisebox{.5pt}{\textcircled{\raisebox{-.9pt} {1}}}, \raisebox{.5pt}{\textcircled{\raisebox{-.9pt} {2}}} and \raisebox{.5pt}{\textcircled{\raisebox{-.9pt} {3}}} in the inelastic cotunneling regime. As shown in Fig.~\ref{fig4}, $F$ presents sizable modulations, on the order of unity, depending on the position of the excited states transitions with respect to source and drain electrochemical potentials, that match the regions delimited by the blue dashed lines. This appears clearly when taking line cuts of the data at fixed $\VDS$, as a function of $\Vg$: just above the inelastic cotunneling onset, at $\VDS=2.7$~mV (dark brown line in the bottom panel of Fig.~\ref{fig4}), $F$ is non-monotonous inside the diamond, with a local minimum at $F\approx1.7$ in the region labeled \raisebox{.5pt}{\textcircled{\raisebox{-.9pt} {3}}} in Fig.~\ref{fig3-GandI}. Larger values of $F$ on either side of that minimum can thus be attributed to COSET processes such as \raisebox{.5pt}{\textcircled{\raisebox{-.9pt} {1}}} and \raisebox{.5pt}{\textcircled{\raisebox{-.9pt} {2}}}. As $\VDS$ is increased, all cotunneling processes end with an excited transition in the transport window, and the local minimum vanishes, as illustrated by the line cut at $\VDS=4.8$~mV (brown line in the bottom panel of Fig.~\ref{fig4}). The measured modulations of $F$ in the inelastic cotunneling regime mimic those predicted in \cite{Aghassi2008, Kaasbjerg2015}, indicating that super-Poissonian current fluctuations indeed stem from different mechanisms depending on the position of the excited states transitions with respect to drain and source electrochemical potentials.

\begin{figure}[ht]
\centering
\includegraphics[width=0.45\textwidth]{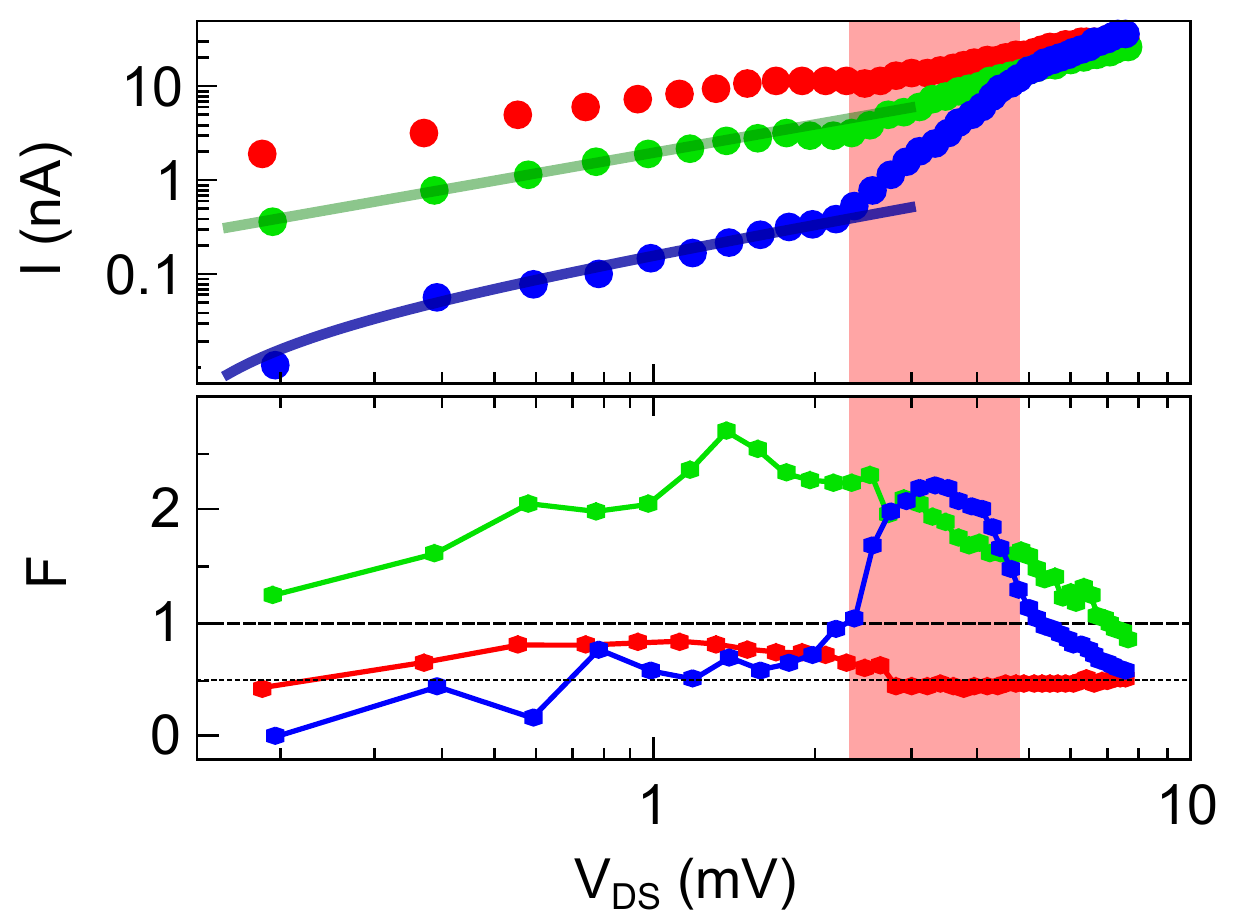}
\caption{\label{fig5}Current $I$ (top panel) and Fano factor $F$ (bottom panel) as a function of $\VDS$ in (resp.) log-log and semi-log scale, for three different values of $\Vg$, located by the  vertical arrows in Fig.~\ref{fig4}: $\Vg=-0.7935~\mathrm{V}$ (red), $\Vg=-0.795~\mathrm{V}$ (green), and $\Vg=-0.8~\mathrm{V}$ (blue). Continuous lines in the top panel: linear fits of the low-$\VDS$ data. The sequential tunneling ($F=0.5$) and Poissonian ($F=1$) values of $F$ are indicated by (resp.) horizontal dotted and dashed lines in the bottom panel. In both panels, the red area indicates the inelastic cotunneling range defined for $\Vg=-0.8~\mathrm{V}$.}
\end{figure}

Fig.~\ref{fig4} shows that, unexpectedly, $F$ takes super-Poissonian values even for $\VDS$ below the inelastic cotunneling onset, close to the edges of the diamonds. This regime of comparatively large fluctuations is quantitatively different from the inelastic cotunneling regime, as illustrated by Fig.~\ref{fig5}, where we have plotted the $\VDS$ dependence of both $I$ and $F$, in (resp.) log-log and semi-log scale, for three values of $\Vg$ leading to different transport regimes. These values are indicated by vertical arrows in Fig.~\ref{fig4}: for $\Vg=-0.7935$~V (red), Coulomb blockade is lifted, and the tunneling through the quantum dot is sequential, leading to comparatively large currents. This regime yields usual, sub-Poissonian values of $F$, saturating at the symmetric value $F=0.5$ at large $\VDS$. The data at $\Vg=-0.8$~V (blue) is similar to the one shown in Fig.~\ref{fig3-GandI}b, first showing a linear $I(\VDS)$ behavior in the elastic cotunelling regime (blue line), followed by an approximative $\left( \VDS \right)^4$ dependence in the inelastic cotunneling regime. As discussed above, this leads to super-Poissonian values of $F$ only in the inelastic cotunneling regime. The data at $\Vg=-0.795$~V (green), close to the edge of the diamond, differs markedly from the previous regimes: indeed, while here $I(\VDS)$ remains essentially linear, taking intermediary values in the nA range, $F$ takes large, super-Poissonian values in the whole range of $\VDS$, in particular well below the inelastic cotunneling onset (appearing as the edge of the red shaded areas in both panels of Fig.~\ref{fig5}). This behavior of comparatively large, linear $I(\VDS)$ together with super-Poissonian fluctuations below the inelastic cotunneling onset appears in both $N-1$ and $N$ diamonds shown in Fig.~\ref{fig4}, for positive and negative $\VDS$ (see also \cite{SM} for maps of the measured current versus $\VDS$ and $\Vg$). Note, however, that 1) it is not observed in all Coulomb diamonds for this device, and 2) surprisingly, it only appears on \textit{one side} of each diamond, namely, towards the $N\leftrightarrow N-1$ degeneracy point of the $N$-hole diamond. This asymmetry is also seen to some extent in the inelastic cotunneling regime~\cite{SM}. To our knowledge, there is no straightforward mechanism leading to enhanced current and fluctuations in the \textit{elastic} cotunneling regime. Among the possible explanations, dynamical channel blockade \cite{Belzig2005} can be ruled out by the fact that it occurs \textit{outside} of the Coulomb diamonds, and that it requires the presence of excited states at energies below $2$ meV, which are not clearly observed in the differential conductance. Another mechanism stems from the possible presence of nearby charge traps poorly coupled to the leads. Single dopants located below the spacers can act as such charge traps \cite{Hofheinz2006a}, randomly switching the conduction state of the quantum dot, thereby increasing current fluctuations. Previous studies on similar nanowire MOSFETs showed that the presence of such traps is characterized in $dI/d\VDS$ measurements by large scale $\Vg$ periodicity and phase shifts in the Coulomb diamonds \cite{Hofheinz2006a, Escott2010,Golovach2011, Villis2014}. While these features do not appear clearly in the measured conductance, similar ones can be seen in the measured shot noise~\cite{SM}; it is thus not entirely unlikely that the presence of one or more charge traps gives rise to the measured enhanced fluctuations at low $\VDS$.

In summary, we have observed for the first time clear modulations of super-Poissonian fluctuations in the inelastic cotunneling regime, which stem from distinct mechanisms depending on the chemical potential of the quantum dot. We have also observed previously unreported enhanced fluctuations in the elastic cotunneling regime, which might be attributed to nearby charge traps.

We warmly thank J. Splettstoesser, M. Misiorny, C. Flindt, Ch. Strunk, A. Donarini, R. Maurand, F. Portier and C. Altimiras for enlightening discussions. This work was funded by the CEA (Programme Transversal Nanosciences \textit{NanoSiN}) and the ERC (ERC-2015-STG \textit{COHEGRAPH}). M. Seo acknowledges support from \textit{Enhanced Eurotalents} programme.

\cleardoublepage



\section*{Supplementary material}

# Supplemental Material for
# Strongly correlated charge transport in silicon MOSFET quantum dots

M. Seo,[1] P. Roulleau,[1] P. Roche,[1] D.C. Glattli,[1] M. Sanquer,[2] X. Jehl,[2] L. Hutin,[3] S. Barraud,[3] and F.D. Parmentier[1]

[1]*SPEC, CEA, CNRS, Université Paris-Saclay, CEA Saclay 91191 Gif-sur-Yvette cedex, France*
[2]*Univ. Grenoble Alpes, CEA, INAC-PHELIQS, 38000 Grenoble, France*
[3]*CEA, LETI, Minatec Campus, 38000 Grenoble, France*
(Dated: 5th June 2018)

## MEASUREMENT SETUP

The measurements were performed in a dry He3 fridge, with a base temperature of 290 mK. Simultaneous measurements of low-frequency conductance and noise of the sample are performed using the circuit shown in Fig. S1. The cryoamps used in the noise measurements are home-made amplifiers based Agilent ATF 34143 HEMTs, with gain ~ 4 and input voltage noise ~ 0.14 nV/$\sqrt{\text{Hz}}$. The two noise measurement lines allow to perform both auto-correlation noise measurement, as well as cross-correlation noise measurement. For each value of $V_\text{g}$, the zero $V_\text{DS}$ noise offset was subtracted to remove the contribution of the thermal noise of the sample and measurement circuit [34]. The shot noise $S_{II}$ discussed in the main text and below corresponds to the excess cross-correlated noise.

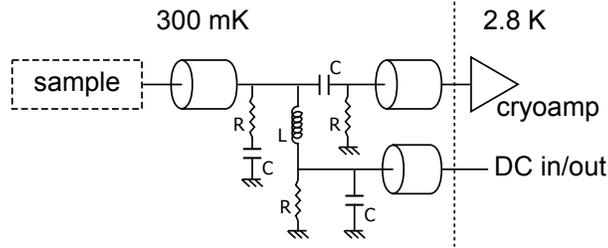

Figure S1. Diagram of the measurement circuit for low-frequency conductance and noise. The values of the different elements in the circuit (corresponding to discrete CMS components) are respectively $R$ = 5 k$\Omega$, $L$ = 22 $\mu$H and $C$ = 22 nF. The total shunt capacitance of the *Lakeshore* coaxial wires between the sample and the cryoamp is about 90 pF. A second, identical circuit is connected to the other lead of the sample, allowing for cross-correlation noise measurements.

The noise measurement setup was carefully calibrated on a regular basis to check the gain stability. The calibration was performed by recording the *RLC* resonance spectra of the two measurement chains at various temperatures between 290 and 600 mK. The elements of the noise measurement circuit where calibrated at low temperature in an independent run without sample. In practice, the gains were found to be constant, with fluctuations between different calibrations and cooldowns below 2 %. The resistors shunting the sample in dc (through the inductor) where calibrated by completely pinching the quantum dot. Their values were found to change by less than 1 % as a function of drain source voltage $V_\text{DS}$. Still this small dependence is taken into account in our measurements of $I(V_\text{DS})$.

## ADDITIONAL DATA

### Main sample - AAD398W1-P02 - Full measurement range

The main sample, the analysis of which is presented in the main text, dubbed AAD398W1-P02, is a p-type device with nominal dimensions for the channel W=30 nm; L=20 nm.

Fig. S2 shows the measurements of the differential conductance $dI/dV_\text{DS}$, current $I$, shot noise $S_{II}$ and Fano facto $F$ as a function of gate voltage $V_\text{g}$ and drain-source voltage $V_\text{DS}$, over the full range of measurements. Due to the large dynamic range of the three first measured quantities, we use log-based functions to plot their variations. For small negative $V_\text{g}$ > −0.65 V, the quantum dot is pinched and its conductance is zero. From the appearance of the first Coulomb diamond, we estimate the number of charges in the dot. The differential conductance measurements show that sweeping the gate voltage towards larger negative values not only increases the number of charges in the dot, but also changes the couplings to source and drain (the resonances become larger, and broader), and modifies the geometry





















































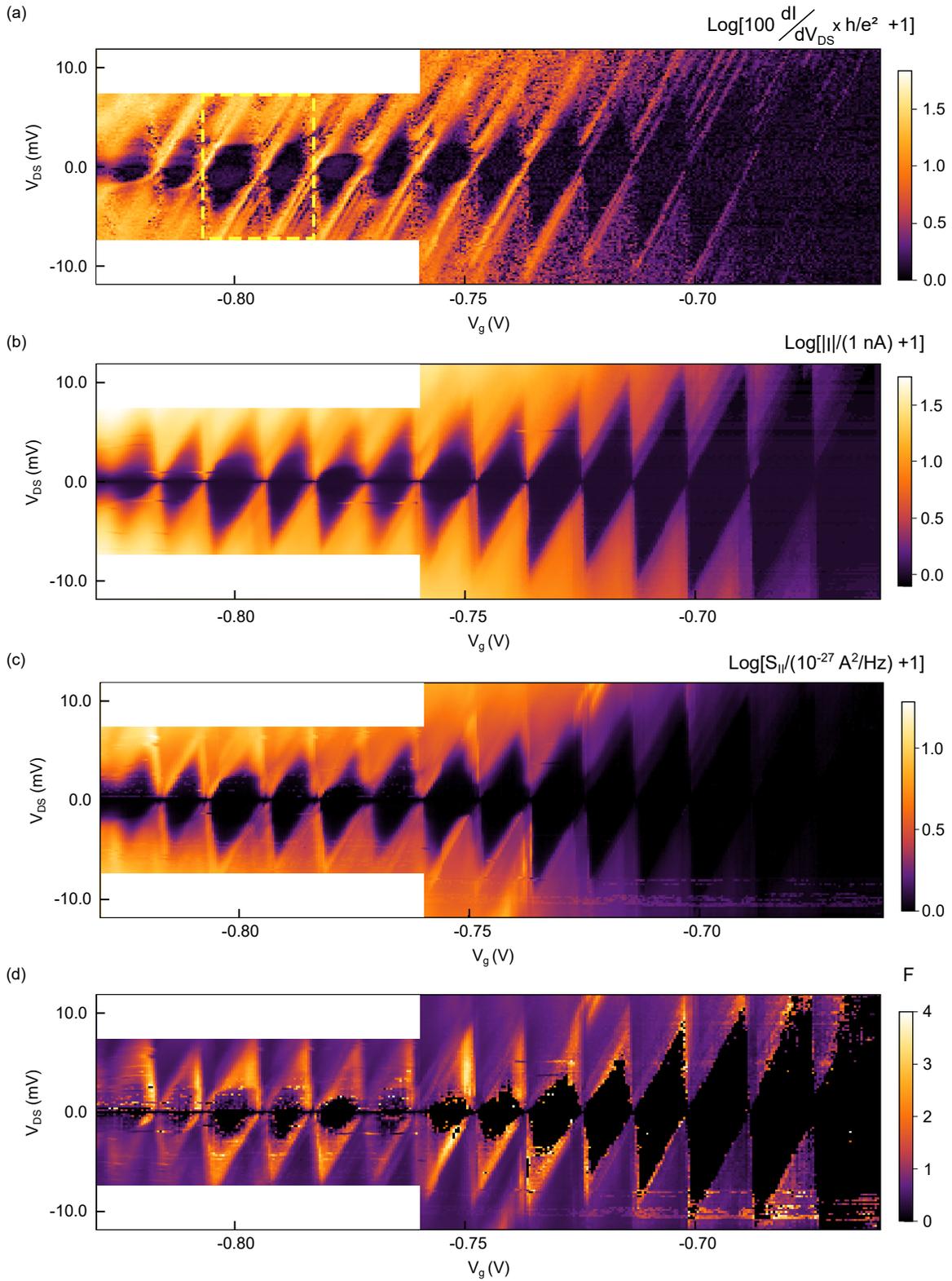

Figure S2. Measurements of differential conductance, current, shot noise and Fano factor of the full range of $V_g$ and $V_{DS}$. **(a)**, Plot of $\text{Log}\left[100 \times dI/dV_{DS}\frac{h}{e^2} + 1\right]$ versus $V_{DS}$ and $V_g$. The measurement range presented in the main text is delimited by the yellow dashed square around $V_g = -0.8$ V. **(b)**, Plot of $\text{Log}\left[|I|/(1\text{ nA}) + 1\right]$ versus $V_{DS}$ and $V_g$. **(c)**, Plot of $\text{Log}\left[S_{II}/(10^{-27}\text{ A}^2/\text{Hz}) + 1\right]$ versus $V_{DS}$ and $V_g$. **(d)**, Plot of $F$ versus $V_{DS}$ and $V_g$. The Log-based functions in **(a)**, **(b)** and **(c)** allow visualizing the full range of differential conductance, current and noise.

of the dot. This can be seen by the fact that the Coulomb diamonds become gradually smaller and tilted towards more even capacitive couplings between gate and dot and between drain and dot. Around $V_g = -0.74$ V, first signatures of inelastic cotunneling appear, along with super-Poissonian noise. In this regime several diamonds present features similar as those of the two diamonds discussed in the main text, shown by the yelow dashed rectangle in Fig. S2a. As mentioned in the main text, super-Poissonian values of $F$ are observed chiefly on the right side of each diamond (i.e. for smaller negative $V_g$). Superpoissonian noise is also observed at very low energy in the diamonds situated left of the ones analyzed in the main text, that is for a more open quantum dot. It is also worth noting that the shot noise $S_{II}$ present features reminiscing of the conductance of a quantum dot in the presence of an additional dopant [27, 39], i.e. slightly shifted, superimposed diamonds, resulting of an apparent doubling of the charge degeneracy point. This could provide an explanation to the enhanced fluctuations observed at low energy in the vicinity of the charge degeneracy point in the diamonds analyzed in the main text.

### Main sample - AAD398W1-P02 - Additional data for the range presented in the main text

The conductance measurements shown in the main text allow us to extract the lever arms for gate and drain in the gate voltage range analyzed in the main text: gate lever arm $\eta_g \approx 0.69$ meV/mV, and drain lever arm $\eta_D \approx 0.091$ meV/mV. We show in Fig. S3 the measurement in of the current in that range, in linear scale (Fig. S3a), and logarithmic scale (Fig. S3b). The current shows Coulomb diamond-like features accurately matched by the resonance lines extracted from the conductance measurements. Note that since the two measurements were taken a few days apart, a very slight shift in gate voltage occurred. This shift, smaller than the minimal increment in gate voltage $\delta V_g = 0.5$ mV, was compensated by shifting the current data by 0.25 mV. We have checked that the results, particularly the Fano factor, do not change qualitatively whether this shift is either not compensated, or over-compensated by shifting the current data by a whole $\delta V_g$. Steps in the current data were furthermore used to extract the excited transitions lines resonant with the source (i.e. with negative slope), that do not appear clearly in the $dI/dV_{DS}$ measurements. These extracted lines perfectly match with resonance features in the conductance in which a positive-slope resonance line abruptly vanishes as it crosses a negative-slope resonance line (see main text Fig. 3a).

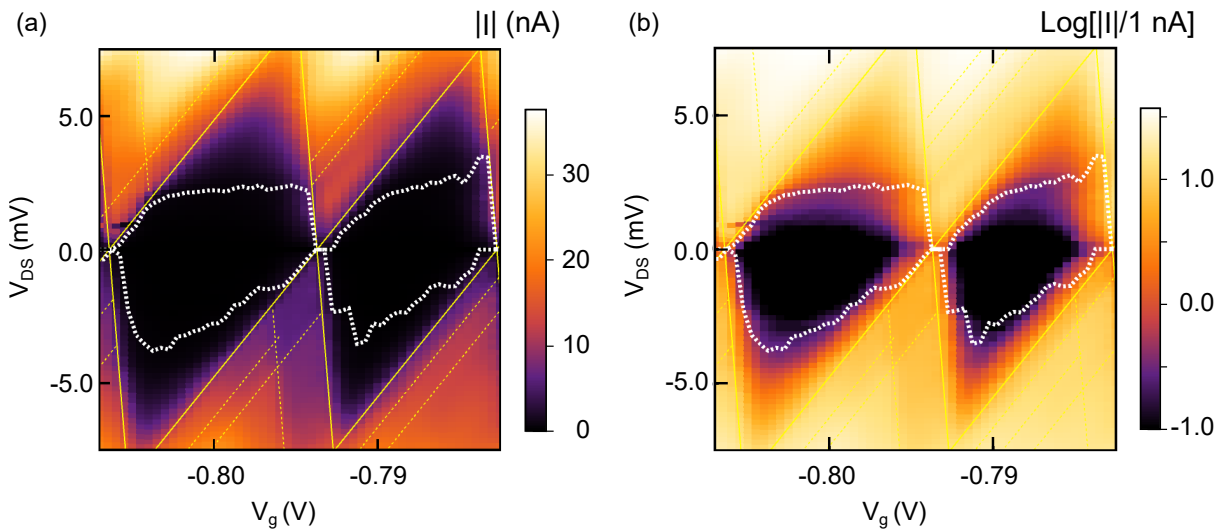

Figure S3. Measurement of the current in the range shown in the main text. (a), Current versus $V_{DS}$ and $V_g$ in linear scale. The yellow and white lines are the same as those discussed in the main text. (b), Same data, in log scale: plot of Log$[|I|/(1\ \text{nA})]$ versus $V_{DS}$ and $V_g$.

Plotting the current in log scale emphasize the cotunneling features, showing that $I$ is markedly larger beyond the inelastic cotunneling threshold (white dashed line in Fig. S3, see main text for details). However, modulations in its amplitude due to COSET are not clearly visible, beyond an obvious increase close to the edges of the diamonds. Importantly, Fig. S3b shows that in the elastic cotunneling regions where super-Poissonian values of $F$ were observed (see main text Fig. 4), $I$ also takes large values, comparable to the values in the inelastic cotunneling regime. We

emphasize the fact that despite its comparatively large amplitude, $I(V_{DS})$ remains linear in this low-energy regime.

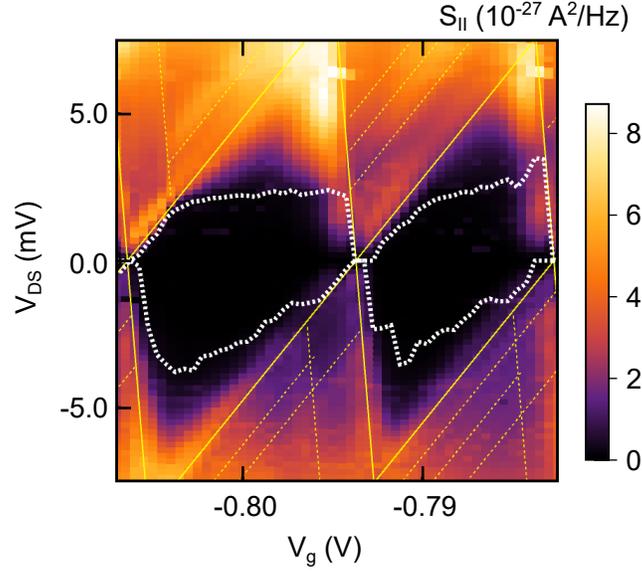

Figure S4. Measurement of $S_{II}$ versus $V_{DS}$ and $V_g$ in the range shown in the main text. The yellow and white lines are the same as those discussed in the main text.

Fig. S4 shows the measurements of the shot noise $S_{II}$ in the same gate voltage range. $S_{II}$ also presents large values beyond the inelastic cotunneling onset, as well as in the low-energy, elastic cotunneling regime mentioned above. Note that outside of the diamonds, the shot noise presents modulations well matched by the excited resonance lines as well as by the features shown in the current measurements.

$I(V_{DS})$ power laws

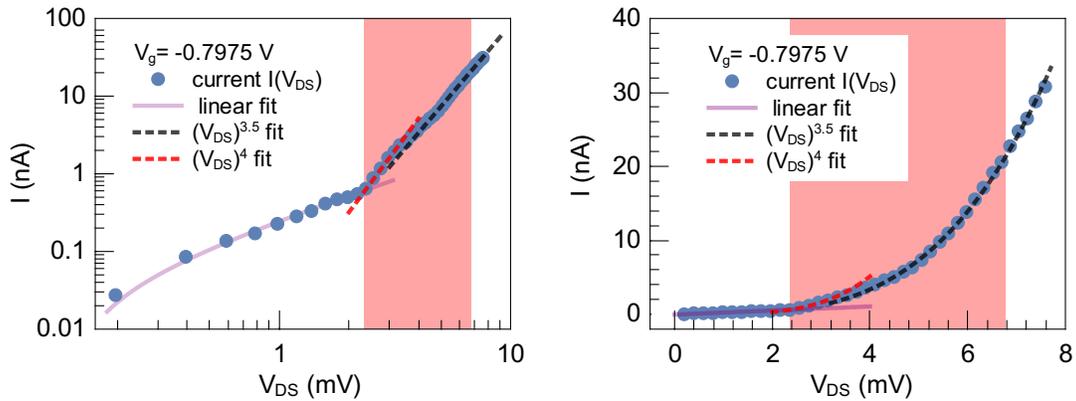

Figure S5. $I(V_{DS})$ characteristics (for positive $V_{DS}$) at $V_g = -0.7975$ V (same experimental dataset as in main text Fig. 3b). Left: log-log scale, right: linear scale. Continuous violet line: linear fit for $V_{DS} < 2.5$ mV; red dashed line: $(V_{DS})^4$ fit for $2 < V_{DS} < 4$ mV; black dashed line: $(V_{DS})^{3.5}$ fit for $V_{DS} > 3$ mV. In both panel the red shaded area corresponds to the inelastic cotunneling regime.

The onset of inelastic cotunneling is extracted from the change of $I(V_{DS})$ from a linear to a power law behavior [35, 36]. Fig. S5 illustrates, for the same dataset shown in the main text, the fits of $I(V_{DS})$ at $V_g = -0.7975$ V in the various regimes. At low energy ($V_{DS} < 2.5$ mV), corresponding to the elastic cotunneling regime, the experimental data is well reproduce by a linear fit $I[\text{nA}] = -0.276 \times V_{DS}[\text{mV}] - 0.033$ (violet continuous line). Note that because of this small 33 pA offset, probably stemming from our resolution on the calibration of the shunting resistors (see above),

the line in log-log scale is not straight. At $V_{DS} \approx 2.3$ mV, the behavior abruptly changes to a power law; however, precisely extracting the exponent is not straightforward, as it is not constant, slightly changing over ranges in $V_{DS}$ much smaller than a decade. To illustrate this, we plot as a red dashed line a $(V_{DS})^4$ adjustment of the data for intermediary values of $V_{DS}$ (between 2 and 4 mV), and as a black dashed line a $(V_{DS})^3.5$ adjustment for $V_{DS} > 3$ mV. Whatever the precise value of the exponent (as long as it is markedly larger than 1), we use the power law behavior to characterize the onset of inelastic cotunneling for each value of $V_g$. Fig. S6 shows the power law behaviors displayed by the $I(V_{DS})$ characteristics for several gate voltages, corresponding to the right half of the left Coulomb diamond studied in the main text (for $V_g$ ranging from −0.793 V (red line) to −0.802 V (blue line). For $V_{DS}$ below ≈ 2.5 mV, all curves display a linear characteristic, with close to two orders of magnitude in range of amplitudes. The low-slope curves, corresponding to values of gate voltages in the center of the Coulomb diamond, then take power laws with a wide range of exponents reaching $(V_{DS})^7$ for $V_g$ = −0.82 mV (blue curve). Again, from this clear change of behavior we extract the inelastic cotunneling onset. The large-slope curves, correspond to data close to the charge degeneracy point, maintain a linear behavior, with some changes of slope, over the whole span of $V_{DS}$.

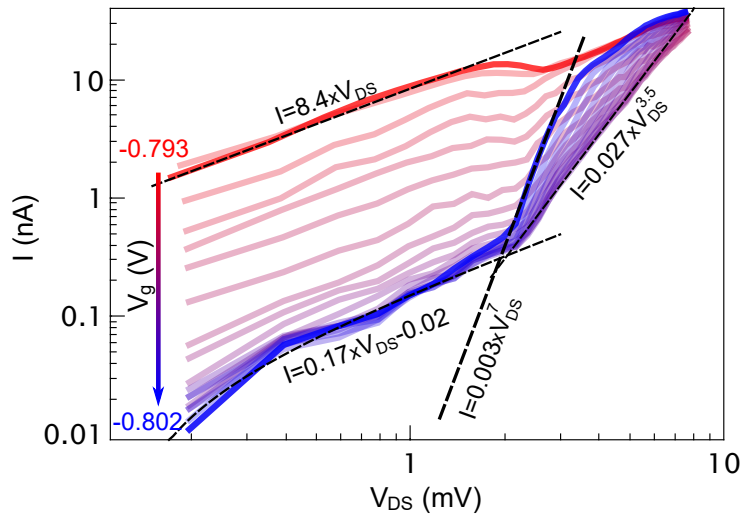

Figure S6. $I(V_{DS})$ characteristics (for positive $V_{DS}$) in log-log scale, for $V_g$ ranging from −0.793 V (red line) to −0.802 V (blue line). The black dashed lines indicate various power laws extracted from the data. Note that the corresponding formulas are expressed in nA and mV.

**Main sample - AAD398W1-P02 - Analysis for other diamonds**

We present here the same analysis, described in the main text, applied to a pair of Coulomb dimaonds measured for larger negative $V_g$ (Fig. S7), corresponding to a more open quantum dot, and 2 pairs of diamonds measured for smaller negative $V_g$ (Fig. S8 and S9), corresponding to a slightly more pinched quantum dot.

Fig. S7 shows differential conductance (**a**), current (**b**) and Fano factor (**c**) measurements for $V_g$ ranging between −0.804 and −0.807 mV. As in the main text, the continuous and dashed yellow lines map the resonances in $dI/dV_{DS}$ corresponding to (respectively) ground and excited transitions. The white dashed lines represents the onset of inelastic cotunneling, extracted from the $I(V_{DS})$ line cuts. In this range, the dot is more open, and the differential conductance and current are slightly larger than in the main dataset, giving rise to pronounced inelastic cotunneling features. In particular, $dI/dV_{DS}$ is clearly non-zero beyond the inelastic cotunneling onset. This dataset presents similar features as the main one in the inelastic cotunneling regime (although less marked); furthermore, it also shows largely enhanced current fluctuations at low $V_{DS}$ in the elastic cotunneling regime, close to the $V_g \to 0$ side of each diamond. This shows the same phenomenology as the super-Poissonian noise measured in the elastic cotunneling discussed in the main text.

Fig. S8 shows differential conductance, current and Fano factor measurements for $V_g$ ranging between −0.724 and −0.749 mV. While $dI/dV_{DS}$ and $|I|$ are smaller by typically a factor 2 compared to the data shown in the main text, $F$ shows a similar behavior, with comparable amplitudes. Indeed, $F$ is clearly larger than 1 in the inelastic cotunneling regime, and drops to values close to 0.5 in the sequential tunneling regime. Modulations of $F$ in the

inelastic cotunneling regime can, here also, be associated with COSET / non-COSET processes discussed in the main text. Note that in this dataset the inelastic cotunneling onset does not, even more markedly than in the main dataset, follow a straight line inside the Coulomb diamond. This non-canonical behavior remains to be understood. Contrarily to the main dataset, and the diamonds located at larger negative $V_g$, the one shown in Fig. S8 does not present super-Poissonian noise below the inelastic cotunneling onset (see also Fig. S2).

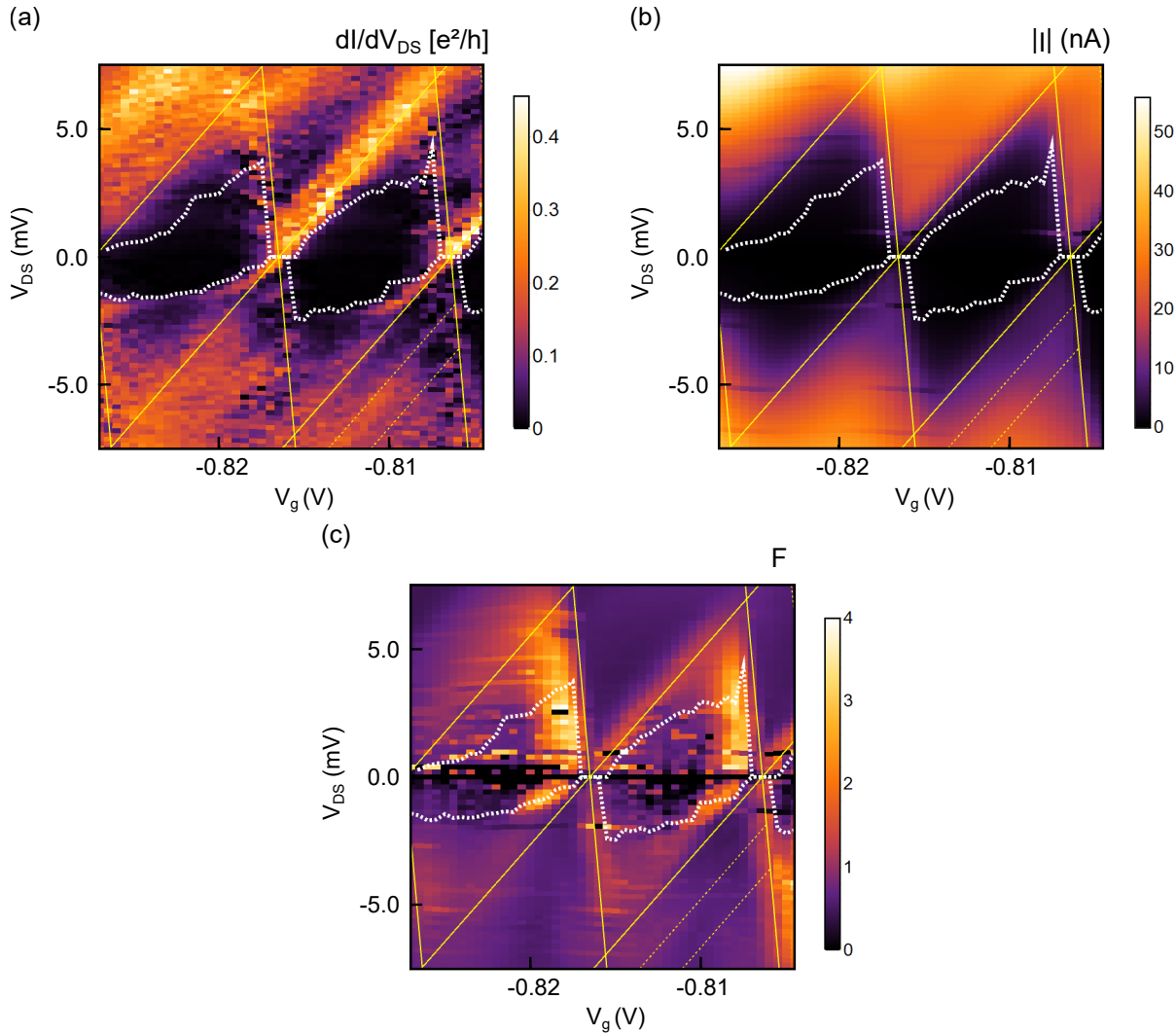

Figure S7. Analysis of diamonds in a slightly more pinched regime. Measurements of (a) $dI/dV_{DS}$, (b) $|I|$ and (c) $F$ versus $V_{DS}$ and $V_g$, for $V_g$ ranging between $-0.724$ and $-0.749$ mV. The yellow and white lines appearing in the three plots correspond to, respectively, the resonance lines extracted from the differential conductance measurements, and the inelastic cotunneling onset extracted from the current measurements.

Fig. S9 shows the application of our analysis on the next pair of diamonds, for $V_g$ ranging between $-0.700$ and $-0.726$ mV. Here the quantum dot becomes markedly more pinched, withe conductances an order of magnitude smaller than in the main text. In that case there is almost no inelastic cotunneling, with the extracted inelastic cotunneling onset (white dashed line) closely following the edges of the diamonds. Nonetheless the measured Fano factor still takes super-Poissonian values beyond the onset, especially for the left diamond centered on $V_g = -0.72$ V. Interestingly, in this case large Fano factor are predominantly found on the *left* side of the diamond, that is towards larger negative $V_g$. Another note-worthy feature of this dataset is the super-Poissonian $F$ measured in the sequential tunneling regime, along an excited resonance line at positive $V_{DS}$, around $V_g = -0.725$ V (see also Fig. S2). This enhancement in the fluctuations probably stems from a blocking state in the transport windows [6, 13–15, 17]; however, no clear feature in either differential conductance or current allows confirming this hypothesis.

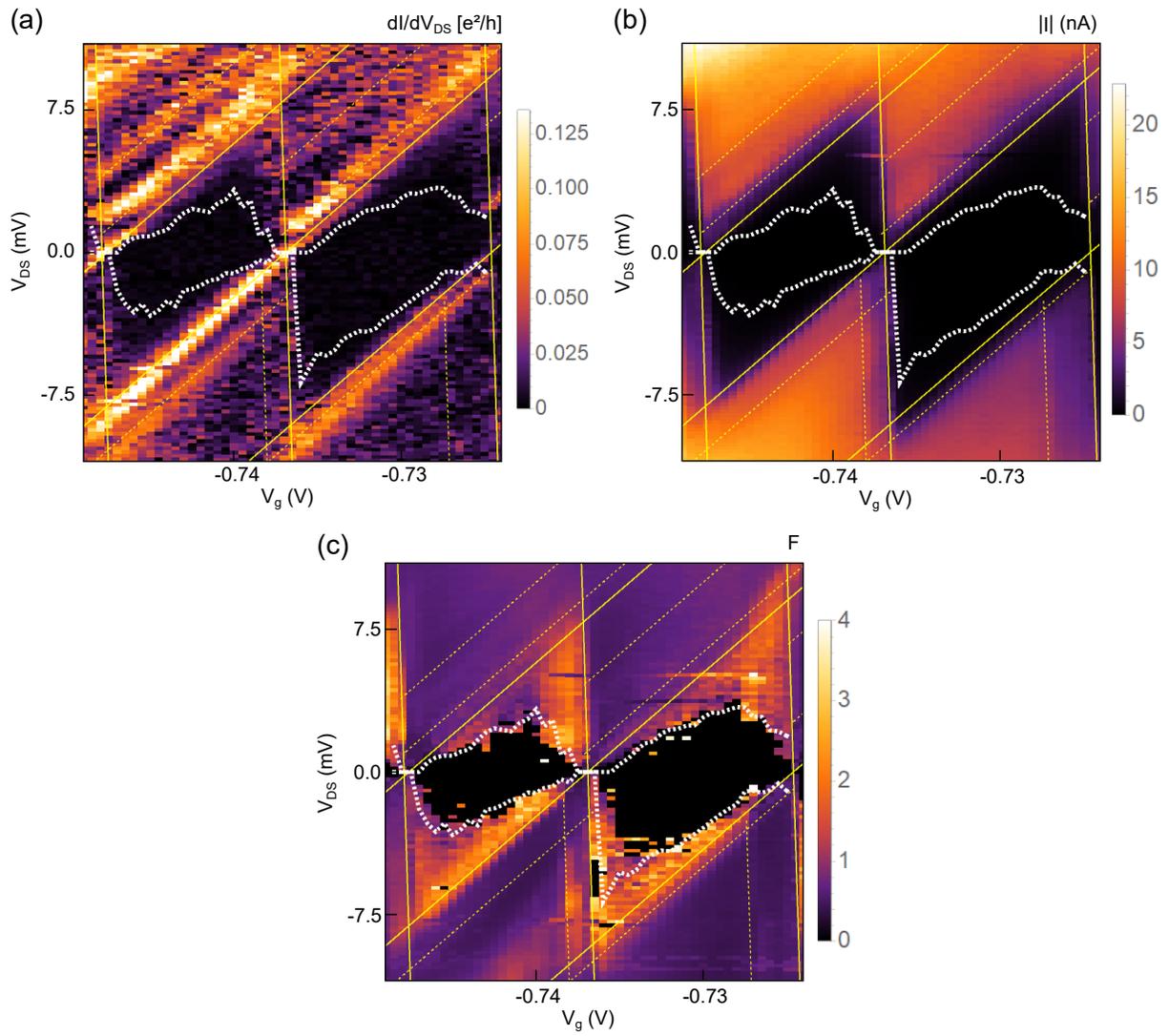

Figure S8. Analysis of diamonds in a slightly more pinched regime. Measurements of (a) $dI/dV_{DS}$, (b) $|I|$ and (c) $F$ versus $V_{DS}$ and $V_g$, for $V_g$ ranging between $-0.724$ and $-0.749$ mV. The yellow and white lines appearing in the three plots correspond to, respectively, the resonance lines extracted from the differential conductance measurements, and the inelastic cotunneling onset extracted from the current measurements.

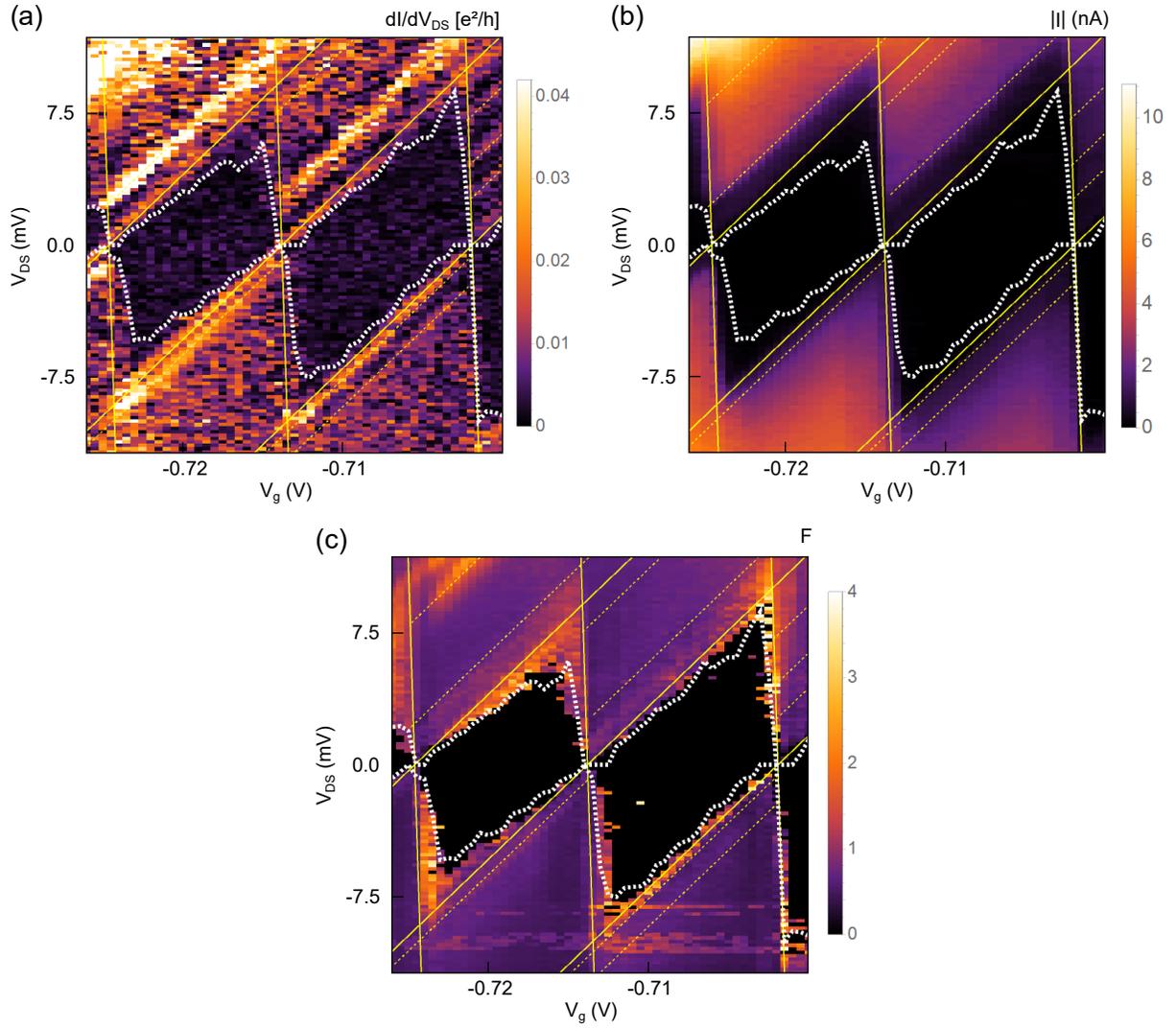

Figure S9. Analysis of diamonds in a slightly more pinched regime. Measurements of (a) $dI/dV_{DS}$, (b) $|I|$ and (c) $F$ versus $V_{DS}$ and $V_g$, for $V_g$ ranging between $-0.700$ and $-0.726$ mV. The yellow and white lines appearing in the three plots correspond to, respectively, the resonance lines extracted from the differential conductance measurements, and the inelastic cotunneling onset extracted from the current measurements.

...

## Additional samples

We now present conductance measurements on two additional devices (AAD398W1-P03 and AAD398W1-P04), all of them p-type, with similar nominal dimensions as the main sample. The results are shown in Fig. S10 and Fig. S11. Both sample present modulations in the differential conductance that are reminiscent of the effects of single dopant close to the quantum dot. [27, 39, 40]. However those two devices were not as stable as the main one (which might be related to the single dopant), preventing us to obtain reliable shot noise data. Note nonetheless that these conductance data, here taken on a much larger range of gate voltage, shows Coulomb blockade properties very similar (in terms of energy scales, and lever arms) as the main sample. As $V_g$ is pushed lowards larger negative values, the quantum dot becomes more and more coupled to the leads, and the conductance increases. Simultaneously, the size of the Coulomb diamonds decreases drastically, reflecting the fact that the quantum becomes larger and larger [26].

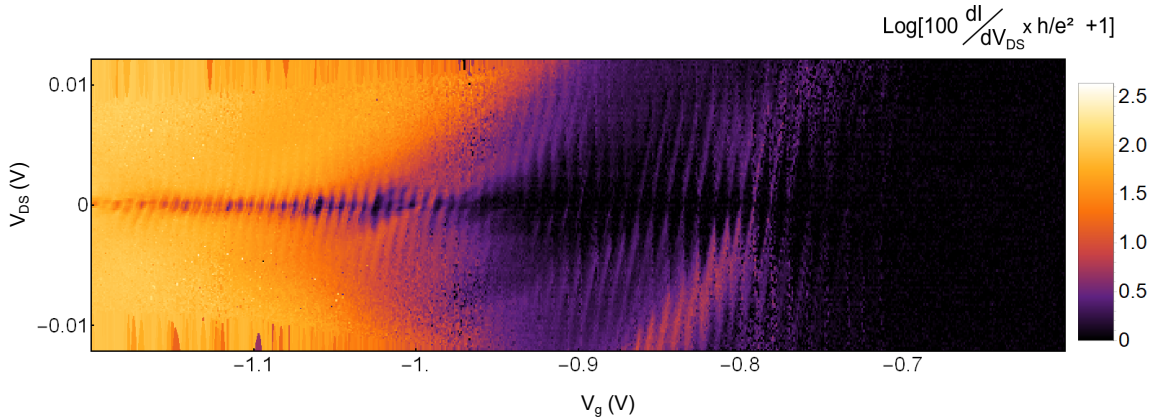

Figure S10. Plot of $\text{Log}\left[100 \times dI/dV_{\text{DS}} \frac{h}{e^2} + 1\right]$ versus $V_{\text{DS}}$ and $V_g$ for sample AAD398W1-P03. The Log-based function allows visualizing the full range of differential conductance.

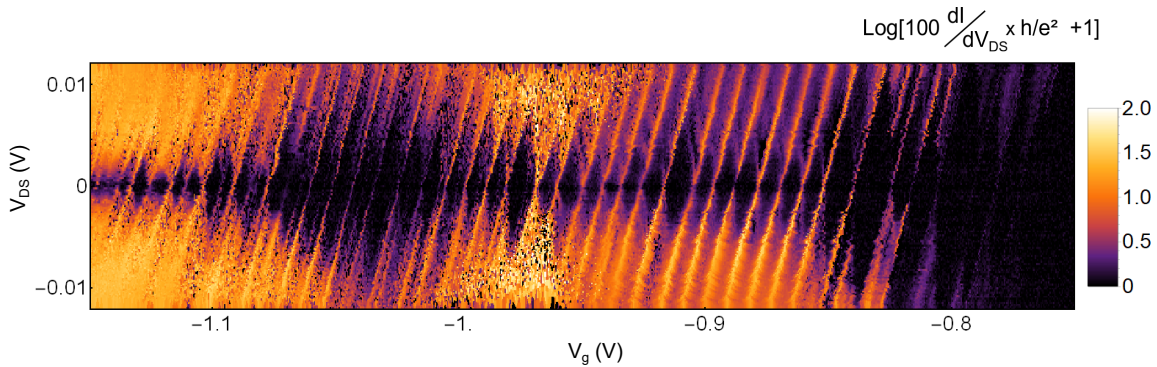

Figure S11. Plot of $\text{Log}\left[100 \times dI/dV_{\text{DS}} \frac{h}{e^2} + 1\right]$ versus $V_{\text{DS}}$ and $V_g$ for sample AAD398W1-P04. The Log-based function allows visualizing the full range of differential conductance.



\end{document}